\documentclass[authoryear,5p]{elsarticle}

\usepackage{graphicx}   
\usepackage{balance}    
\usepackage{url}
\usepackage{subfig}

\usepackage{amssymb}
\usepackage{amsmath}
\usepackage{graphicx}
\usepackage{url}
\usepackage{tabularx}
\usepackage{amssymb}
\usepackage{booktabs}
\usepackage{caption}
\usepackage{lineno}

\DeclareCaptionFormat{subfig}{\figurename~#1#2#3}
\DeclareCaptionSubType*{figure}
\journal{Computers in Human Behavior (forthcoming)}

\begin{document}
\begin{frontmatter}

\title{Does Offline Political Segregation  Affect the Filter Bubble? An Empirical Analysis of Information Diversity for Dutch and Turkish Twitter Users}

\author[values]{E. ~Bozdag\corref{cor1}}
\ead{v.e.bozdag@tudelft.nl}

\author[philips]{Q. ~Gao}
\ead{q.gao@philips.com}

\author[wis]{G.J. ~Houben}
\ead{g.j.p.m.houben@tudelft.nl}

\author[system]{M.E. ~Warnier}
\ead{m.e.warnier@tudelft.nl}

\cortext[cor1]{Corresponding author}
\address[values]{Delft University of Technology, Department Values and Technology, P.O. Box 5015 2600 GA Delft, The Netherlands}
\address[philips]{Philips Research, Eindhoven, the Netherlands}
\address[wis]{Delft University of Technology, Web Information Systems Group, Delft, the Netherlands }
\address[system]{ Delft University of Technology, Section of System Engineering. Delft, the Netherlands}

%

\begin{abstract}
From a liberal perspective, pluralism and viewpoint diversity are seen as a necessary condition for a well-functioning democracy. Recently, there have been claims that viewpoint diversity is diminishing in online social networks, putting users in a ``bubble", where they receive political information which they agree with. The contributions from our investigations are fivefold: (1) we introduce different dimensions of the highly complex value viewpoint diversity using political theory; (2) we provide an overview of the metrics used in the literature of viewpoint diversity analysis; (3) we operationalize new metrics using the theory and  provide a framework to analyze viewpoint diversity in Twitter for different political cultures; (4) we share our results for a case study on minorities we performed for Turkish and Dutch Twitter users; (5) we show that minority users cannot reach a large percentage of Turkish Twitter users. With the last of these contributions, using theory from communication scholars and philosophers, we show how minority access is missing from the typical dimensions of viewpoint diversity studied by computer scientists and the impact it has on viewpoint diversity analysis.
\end{abstract}

\begin{keyword}
Twitter \sep  diversity \sep polarization \sep politics \sep Turkey \sep Netherlands 
 \end{keyword}

\end{frontmatter}

\section{Introduction}
It is well known that traditional media have a bias in selecting what to report and in choosing a perspective on a particular topic. Individual factors such as personal judgment can play a role during the selection of news for a newspaper. Selection bias,  organizational factors, advertiser and government influences can all affect which items will become news \citep{Bozdag2013}. About 37\% of Americans  see a great deal of political bias in news coverage and 68\% percent  prefer to get political news from sources that have no particular point of view \citep{PewResearch2012}.  Similarly, in a survey performed before the general elections in the UK, 96\% of the population  said they believe they have seen clear bias within the UK media \citep{Wei2013}.  Evidence of bias ranges from the topic choice of the New York Times to the choice of think-tanks that the media refer to \citep{DellaVigna2007}.

Many democracy theorists claim that modern deliberative democracy requires  citizens to have socially validated and justifiable preferences. 
Citizens must be exposed to opposed preferences and viewpoints and should be able to defend their views \citep{Held2006,Offe1990,Dryzek1994}.  Exposure to biased news information can foster  intolerance to opposing viewpoints,  lead to ideological segregation and antagonisms in major political and social issues \citep{Glynn2004,An2012,Saez-Trumper2013}. Being aware of and overcoming bias in news reporting is essential for a fair society, as media has the power to shape voting behavior \citep{Saez-Trumper2013}.

Social information streams, i.e., status updates from social networking sites, have emerged as a popular means of information sharing. Political discussions on these platforms are becoming an increasingly relevant source of political information, often also used as a source of quotes for media outlets ~\citep{Jurgens2011}.  Traditional media are declining in their gatekeeping role to determine the agenda and select which issues and viewpoints reach their audiences \citep{Bruns2011}. Internet users have moved from scanning traditional media such as newspapers and television to using the Internet, in particular social networking sites \citep{An2012}. Social networking sites are thus now acting as gatekeepers \citep{Bozdag2013}. 

Communication theorists argue that the traditional media are declining in their gatekeeping role to determine what is ``newsworthy" and select which issues and viewpoints will reach their audiences \cite{Bruns2011}. It is often argued that the Internet, by promoting equal access to diverging preferences and opinions in society, actually increases information diversity.  Many scholars characterize the online media landscape as the ``age of plentyÓ, with an almost infinite choice and unparalleled pluralization of voices that have access to the public sphere \citep{Karppinen2009}. Some argue that social media will  disrupt the traditional elite control of media and amplify the political voice of non-elites and minorities \citep{Castells2011}.
Still others  claim that tools such as Twitter are neutral spaces for collaborative news coverage operated by third parties outside the journalism industry. As a result, the information curated through collaborative action on such social media platforms should be expected to be drawn from a diverse, multi-perspectival range of sources \citep{Bruns2011}. Some further claim that platforms such as Twitter are neutral communication spaces, and offer a unique environment in which journalists are free to communicate virtually anything to anyone, beyond many of the natural constraints posed by organizational norms that are existing in traditional media \citep{Lasorsa2012a}.

On the other hand, there are skeptical voices that  argue that the Internet has not fundamentally changed the concentrated structure typical of mass media, but reflects the previously recognized inequalities \citep{Karppinen2009}. It is also argued that it has   brought about new forms of exclusion and hierarchy \citep{Suoranta2009}. While it has increased some sort of political participation, it has empowered a small set of elites and they still strongly shape how political material is presented and accessed \citep{Hindman2008}. Others have pointed out the danger of ``cyberbalkanization"" caused by the Internet\citep{Sunstein2002, Pariser2011c}. They argue that the filters we choose on the Internet, or the filters that are imposed upon us will weaken the democratic process. This is because it  will allow citizens to join into  groups  that share their own views and values, and cut themselves off from any information that might challenge their beliefs. Group deliberation among like-minded people can create polarization; individuals may lead each other in the direction of error and falsehood, simply because of the limited argument pool and the operation of social influences. 

It is thus very important to verify whether viewpoint diversity is diminishing in social media and whether cyberbalkanization indeed occurs. There are empirical studies that have observed a high level of information diversity in Twitter and Facebook, mainly due to retweets and weak-ties \citep{Bakshy2012,An2011,Sun2013}. While being very valuable contributions to the literature, these studies often focus on American users and they define information diversity either as ``novelty", or ``source diversity". However, as we will show below, novel information does not necessarily contribute to information diversity and highly competitive media markets  with many sources may still result in excessive sameness of media contents.  As we will argue, marginalized members of segregated groups, structurally underprivileged actors and minorities must receive  special attention and just measuring number of available sources will not guarantee viewpoint diversity.

In this paper, we contribute with a framework to analyze and understand the impact of political culture in Twitter.  Rather than reducing the concept viewpoint diversity to a single quantity or metric, we introduce different dimensions of viewpoint diversity, based on previous studies and the theory from communication studies and political philosophy. 
In addition, we provide a set of new metrics and operationalize them. Finally, we present the result of a case study we performed for Dutch and Turkish Twitter users using this framework. We show that minority users cannot reach a large percentage of the studied Turkish Twitter users and political culture is making a difference.

 \section{Empirical Studies of Information Diversity in Social Media}
An empirical study performed by Facebook suggests that online social networks may  increase the spread of novel information and of diverse viewpoints. According to Bakshy (2012), even though people are more likely to consume and share information that comes from close contacts that they interact with frequently, the vast majority of information comes from contacts that they interact with infrequently. These so-called ``weak-ties"  \citep{Granovetter1981} are also more likely to share novel information. However, there are some concerns with this study. First, Facebook does not provide open access to everyone, thus we can not repeat or reproduce the results using Facebook data. Second, our weak ties give us access to new stories that we would not otherwise have seen, but these stories might not be different ideologically from our own general worldview. They might be novel information, but not particularly diverse.  The concepts serendipity, diversity and novelty are different from each other \citep{Sun2013}. The Facebook research does not indicate whether we encounter and engage with news that opposes our own beliefs through ``weak-links". 

Twitter, with its API, provides an excellent environment for information diversity research. An et al. \citeyearpar{An2012} observe extreme polarization among media sources in Twitter. In another study, they found that, when direct subscription is considered alone, most Twitter users receive only biased political views they agree with \citep{An2011}. However, they note that the news media landscape changes dramatically under the influence of retweets, broadening the opportunity for users to receive updates from politically diverse media outlets. Sun et al. \citeyearpar{Sun2013} performed an empirical study using statistical models to identify serendipity in Twitter and Weibo. Using likelihood ratio test and by measuring unexpectedness and relevance, they observe high levels of serendipity in information diffusion in microblogging communities. Saez-Trumper et al. \citeyearpar{Saez-Trumper2013} found that political bias is evident in social media, in terms of the distribution of tweets that different stories receive. Further, statement bias is evident in social media; a more opinionated and negative language is used than the one used in traditional media. Twitter users are  more interested in what is happening  directly around them and what is happening to those around them. While communities talk about a broad range of news, Twitter users dedicate most of their tweets to a few of them \citep{Saez-Trumper2013}. Wei et al. \citeyearpar{Wei2013} found that individual journalists have the strongest influence on Twitter for UK users. Further, they observed that all influential British Twitter users (mainstream media, journalists and celebrities)  display some kind of bias towards a particular political party in their tweets. Jurgens et al. \citeyearpar{Jurgens2011} show that certain individual German Twitter users act as gatekeepers, especially in the distribution of political information. Those users are also not neutral hubs. They tend to curate political information and post messages that they find important \citep{Jurgens2011}.

\section{Theory}
In this section, we first give a short overview  ``information diversity" and explain why it is a vital value for a democratic society. Later, we show different dimensions of this value and show how it can be defined. 

\subsection{Information Diversity}
A cyberbalkanized Internet or ``filter bubble" is not acceptable in different models of modern democracy.  Aggregative versions of democracy hold that legitimacy lies in the fair counting of votes casted by informed voters  \cite{Held2006}.  Deliberative democrats  on the other hand hold that a decision is only legitimate if it is determined by fair, informed deliberations \citep{Fishkin1993,Cohen2009}.  Because no set of values or preferences can claim to be correct by themselves, they must be justified and tested through social encounters which take the point of view of others into account \citep{Held2006}.  In addition to the normative value of discussion, information-sharing is required for many of the practical benefits that proponents of deliberation hope deliberative institutions will provide, such as higher quality policy, greater appreciation of the views of the opposing side, cultural pluralism and citizen welfare \citep{Napoli1999}. According to deliberative democrats, we must focus on why and how we come to adopt our views, and whether they can be defended in a complex social setting with people with opposed preferences. This will complement voting, the necessary mode of participation, by a ``conscious confrontation of one's own point of view with an opposing point of view, or of the multiplicity of diverse viewpoints that the citizen, upon reflection, is likely to discover within his or her own self"\citep{Offe1990}. Under conditions of ideal deliberation, `no force except that of the better argument is exercised' \citep{Habermas1975}. 
 
Information diversity is also an important concept in communication studies. The freedom of media, a multiplicity of opinions and the good of society are inextricably connected \citep{Napoli1999}. \emph{Free Press theory}, a theory of media diversity, states that we establish and preserve conditions that provide many alternative voices, regardless of intrinsic merit or truth,  with the condition that they emerge from those whom society is supposed to benefit its individual members and constituent groups\citep{VanCuilenburg2003}. What is good for the members of the society can only be discovered by the free expression of alternative goals and solutions to problems, often disseminated through media \citep{Napoli1999}.  

While many scholars from different disciplines agree that information diversity is an important value that we should include in the design of institutions, policies and online services, this value is often reduced to a single definition, such as ``source diversity", or ``hearing the opinion of the other side". In the next subsections, we explain that just having a deliberation is not enough, and that a  bias against arguments made by deliberators who are in the minority in terms of their interests in the decision being made can exist. 

\subsection{Dimension of Information Diversity}
\label{sec:dimensions-diversity}
Following Napoli \citeyearpar{Napoli1999}, we may distinguish three different dimensions of information diversity. The first dimension is \textit{source diversity}, which is diversity in terms of outlets (cables and channel owners) or program producers (content owners). 
\textit{Content diversity} consists of diversity in format (program-type), demographic (in terms of racial, ethnic, and gender), and idea-viewpoint (of social, political and cultural perspectives). 
The third dimension, \textit{exposure diversity},  deals with audience reach and whether users have actually consumed a diverse set of items.  

In US media policy, with the ``free marketplace of ideas" theory, it is assumed that increasing source diversity will  increase content diversity and exposure diversity will follow these two. American media policy consequently focuses on source diversity by way of competition and antitrust regulation \citep{VanCuilenburg2002}.  However, whether more media competition (more sources) really brings about more media variety is  highly debated and research addressing this relationship has not provided definitive evidence of a systematic relationship \citep{Napoli1999,VanCuilenburg1999,McQuail1983,Karppinen2013}. Highly competitive media markets may still have low content diversity and media monopolies can produce highly diverse supply of media content \citep{VanCuilenburg2002}.  It has also been argued  that to fulfill the objectives of the marketplace of ideas metaphor, policymakers need to focus on  exposure diversity. So, one should not look at availability of different sources or content, but whether the public consumes a diverse set of items \citep{Napoli1999}.

\subsection{Minorities and Openness}
\label{sec:minorityaccess}

Karppinen \citeyearpar{Karppinen2009} argues that the aim of media diversity should not be the multiplication of genre, sources or markets, but giving voice to different members of the society. We should not see diversity as something that can be measured through the number of organizations or channels or just ``having two parties reach all citizens". Karppinen holds that we should  focus on democratic distribution of communicative power in the public sphere and whether everyone has the chance and resources to get their voices heard. Karppinen argues: ``the key task for media policy from the radical pluralist perspective is to support and enlarge the opportunities for structurally underprivileged actors and to create space for the critical voices and social perspectives excluded from the systematic structures of the market or state bureaucracy"\citep{Karppinen2009}.  If democratic  processes and public policies exclude and marginalize members of segregated groups from political influence to the extent that privileged groups often dominate the public policy process, they will magnify the harms of segregation. These ``minorities" must be politically mobilized and included as equals in a process of discussing issues \citep{Young2002}. 

McQuail and van Cuilenburg \citeyearpar{McQuail1983} propose to assess media diversity by introducing two normative frameworks. The norm of \textit{reflection} checks whether ``media content proportionally reflects differences in politics, religion, culture and social conditions in a more or less proportional wayÓ. The norm of \textit{openness} checks whether media ``provide perfectly \textit{equal access} to their channels for all people and all ideas in societyÓ. If the population preferences were uniformly distributed over society, then satisfying the first condition (reflection) would also satisfy the second condition (equal access). However, this is seldom the case \citep{VanCuilenburg1999}. Often population preferences tend toward the middle and mainstreams. In such cases, the media will not satisfy the openness norm, and the  preferences of the minorities will not reach a larger public. This is undesirable, because ``social change usually begins with minority views and movements ($...$) asymmetric media provision of content may challenge majority preferences and eventually may open up majority preferences for cultural change in one direction or another" \citep{VanCuilenburg1999}. Van Cuilenburg \citeyearpar{VanCuilenburg1999} argues  that the Internet has to be assessed in terms of its ability to give open access to new and creative ideas, opinions and knowledge that the old media do not cover yet. Otherwise it will only be ``more of the same".


\section{Polarization in the Netherlands and Turkey}
Before discussing methods and the results of our empirical study that focused on Dutch and Turkish users, we give a short overview of political diversity for those two countries and explain why they are interesting for a case study on information diversity.
\subsection{the Netherlands}
Pillarization (Dutch: ``verzuiling") is a process  that occurred in the Netherlands and reached its highest point in 1950Õs. During this period, several ideological groups making up  Dutch  society were systematically organized as parallel complexes that were mutually segregated and polarized \citep{vanDoorn1956, Post1985}. As part of this social apartheid dividing the population into subcultures, political parties were used for political mobilization of the ideologically and religiously defined groups and social activities were concentrated within the particular categorical group \citep{Steininger1977}. Few contact  existed between the different groups and internally the groups were tightly organized \citep{Lijphart1968}.  Elites at the `top' level communicated, while the ones at the `bottom' did not. 
Pillarization had an effect on parental choice of an elementary school for children, the voting for political parties and the choice on which daily newspaper to read \citep{Kruijt1962}. People belonging to a pillar retreated into their own organizations and entered into a `voluntary' isolation, because they perceive that values important to them were threatened \citep{Marsman1967}.  

Depillarization (Dutch: ``ontzuiling") started in mid 1960's as a democratization process and pillarization has lost much of its significance since the 1960's as a result of secularization and individualization. Even though depillarization has started, many institutional legacies in present-day Netherlands still reflect its pillarized past, for example in its public broadcasting system or in the school system  \citep{Vink2007}.
The Netherlands continues to be a country of minorities, which may be a main reason that consensus seems so ingrained in the Dutch political culture \citep{vanderHoek2000}. The Dutch parliament currently has 12 political parties. Due to the very low chance of any party gaining power alone, parties often form coalitions.

The Netherlands has created several media policies  to implement diversity in the media. 
The Media Monitor, an independent institution, measures ownership concentration, editorial concentration and audience preferences \citep{Aslama2007}. It also measures diversity of television programming on the basis of a content classification system, by categorizing  program output in categories like news and information, education, drama, sports, etc.  \citep{Mediamonitor,VanCuilenburg2002}. 
\subsection{Turkey}

Turkey has regularly held free and competitive elections since 1946.  The country has alternated between a two-party political system and a multi-party system. Electoral politics has often been dominated by highly ideological rival parties and military inventions changed the political landscape several times \citep{tessler2004}. 
Elections in 2002 led to a two-party parliament, partially due to a ten per cent threshold. The Justice and Development Party\footnote{Adalet ve Kalkinma Partisi} (AKP) won the elections and still is the ruling party, having an absolute majority. The parliament is currently formed by 4 political parties. While AKP has 59\% of the MP's, secular Republican People's Party\footnote{Cumhuriyet Halk Partisi} (CHP) has 24\%. 

AKP's dominance and the despair and sense of marginalization felt by its opponents threaten to create a political polarization. 
Muftuler-Bas and Keyman \citeyearpar{muftuler2012} argue that ``many  other polarizing social and political struggles remain unresolved in Turkey, and mutually antagonistic groups remain unreconciled. This social and political polarization remains potentially explosive and reduces the capacity for social consensus and political compromise". Similarly Unver \citeyearpar[p.2]{Unver2011} claims that ``the society is pushed towards two extremes that are independent of party politics. (...) Competing narratives and `realities' clash with each other so intensely, that the resultant effect is one of alienation and `otherness' within the society."

Some scholars argue that, the top-down imposition of  concepts such as democracy, political parties and parliament as part of westernization efforts is causing the socio-political polarization in Turkey \citep{Altintas2003}. Agirdir \citeyearpar{Agirdir2010} argues that ``the  system does not breed from the diverse interests and demands of the society, but around the values and interests of a party leader and the narrow crew around her". Economic voting behavior, religiosity, and modern versus traditional orientation seem to be the strongest drivers of polarization \citep{yilmaz2012}. Some argue that, after 2011 polarization has increased and reached its highest points in Turkish history \citep{Ozturk2013}. 
The difference of opinion between different clusters about  secularity, tolerance and political change issues  in total contradiction of each other, therefore a danger of absolute social polarization is imminent \citep{Agirdir2010}. Kiris \citeyearpar{Kiris2011}  observes an identity-based polarization, between secularists and islamists, between Turkish nationalists and Kurdish Ethnic Nationalists, and between Alevis and Sunnis (different sects of Islam). 

Turkish Radio Television Supreme Council (RTUK) was established in order to 
control whether Turkish language, Turkish history, historical values, Turkish way of life, thoughts and feelings are being given a significant place in broadcasting programs \citep{Acar2004}. RTUK is sometimes referred as ``the Censure Board" \citep{MuftulerBac2005} and its decisions of penalizing the broadcasters 
have been criticized domestically and internationally \citep{Baris2007,Demir2007}.
RTUK does not have a diversity policy and the lack of diversity in programme-making is said to undermine the quality of the audio-visual media \citep{Baris2007}.

\subsection{Conclusion}
In short, The Netherlands and Turkey are two  different countries in terms of  political landscape and diversity policy. The Dutch society is less polarized than it was half a century ago, while the Turkish society is thought to be heavily polarized. The Dutch Parliament contains many political parties, no party has absolute power to govern alone. Turkey, on the other hand has few political parties represented in the government and the ruling party has almost 60\% of all the seats. Further, the Dutch media is regulated with a diversity policy. While Turkey has a similar institution, it acts more as a censor board and does not employ an active diversity policy. If the social networking platforms mirror the society, then we can expect the Dutch users to receive more diverse content, while the Turkish users to be more polarized and have a less diverse newsfeed. 
\section{Method}
In this section we provide our method of data collection, present our research model and the metrics we operationalized to measure information diversity.
\subsection{Data collection}
In January 2013, over a period of more than one month we crawled microblogging data via the Twitter's REST and  streaming APIs\footnote{https://dev.twitter.com/}. We started from a \textit{seed} set of Dutch and Turkish Twitter users $U_{s}$, who mainly publish news-related tweets. We have selected different types of users including mainstream news media, journalists, individual bloggers and politicians. The list of these ``influential" users were  picked up from different ranking sites. For the Dutch ranking, we used Peerreach\footnote{http://peerreach.com/lists/politics/nl}, Twittergids\footnote{http://twittergids.nl/} and Haagse Twitter-stolp\footnote{http://alleplanten.net/twitter/site/de-resultaten/belangrijke-personen/ }.  For Turkish ranking, we used TwitterTurk\footnote{http://twitturk.com/twituser/users/turk} and TwitterTakip\footnote{http://www.twittertakip.com/}. 

This resulted in two lists for Turkey and the Netherlands, both containing seed users categorized in political groups, which differed per country. We mapped the political leaning of Dutch seed users into five groups and the political leaning of Turkish seed users into nine groups.  We did this using a number of public sources \citep{Krouwel2008, VanderEijk2000,Trendlight2012, Carkoglu2007}. Later, we defined some of the seed users as a ``minority". We did this by selecting seed users who belong to a political party that is either not represented in the parliament, or is represented with few MP's. We also included MP's of a large political party who belong to an ethnic minority. That makes for instance the Kurdish Party BDP and its MP's a minority in Turkey, while we consider the Greens as a minority in the Netherlands. See Appendix A for a list of minorities. Both user groups defined as minorities create about 15\% of the all observed tweets for both countries. 

By monitoring the Twitter streams of $U_{s}$, we were able to add another set of users $U_{n}$, who followed and retweeted at least 5 items from users in $U_{s}$. After removing users who were involved in spam, we had  1981 Dutch users and 1746 Turkish users.  After crawling seed user tweets and identify the retweets made by their followers, we operationalized various metrics. In the following subsections, we explain the translation of research questions into metrics.

\subsection{Research questions}
The main  question in this research is the following: ``Does offline political segregation affect information diversity in Twitter?". To answer this question, we have provided some sub-questions. 
\begin{enumerate}
\item \textbf{Q1- Seed User Interaction:} Do seed  users  from one end of the political spectrum ever tweet links from another category? Do they reply to each other?
The results of this question is relevant to the previously conducted studies that studied media bias on Twitter, such as Wei et al. \citeyearpar{Wei2013}
\item \textbf{Q2 - Source Diversity:} Is the newsfeed of social media users diverse? Are they receiving updates from a diverse set of users?   
Does indirect exposure (e.g., via retweets or weak-links) increase diversity marginally?  Result of these questions are relevant to the previously conducted studies, such as An et al.'s \citeyearpar{An2011}.
\item \textbf{Q3- Output Diversity:} Do users share items  from a diverse set of users or mainly from the same political category? This question is relevant to the framework provided by Napoli, which we have mentioned in Section \ref{sec:dimensions-diversity}.
\item \textbf{Q4 - Openness/Minority Diversity:} Can minorities reach the social media users, so that ``equal access" principle is satisfied? Or can only the "popular"  reach a bigger public?  This question is relevant  to the normative theory of \cite{McQuail1983} and \cite{Karppinen2013}, which we discussed in  Section \ref{sec:minorityaccess}.
\item \textbf{Q5 -  Input-Output Correlation:} Do users post political messages whose political position reflects the political position of those messages that the users receive? Or do the messages they choose to retweet show a political position significantly skewed from  the political position of the messages which they receive? Result of this question is relevant to the previously conducted studies such as Jurgens et al.'s \citeyearpar{Jurgens2011}.

\end{enumerate}

\subsection{Entropy}
While translating the concepts introduced in the previous subsection  into  metrics,  we apply the  following entropy formula  used by van Cuilenburg \citeyearpar{VanCuilenburg2007} to measure  traditional media diversity, which is based on the work of Shannon \citeyearpar{ShannonClaudeE.1948}  :

\begin{equation}
- ( \sum (p_i \log p_i) / log (n) )
\label{eq:entropy}
\end{equation}

\begin{figure}[htbp]
\begin{center}
  \includegraphics[width=0.5\textwidth]{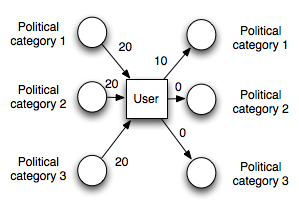}
\caption{Applying entropy}
\label{fig:entropy}
\end{center}
\end{figure}

In \cite{VanCuilenburg2007},  Ò$p_i$ Ò represents the proportion of items of content type category $i$. $n$ represents the number  of content type categories.  We use this formula for calculating source diversity and exposure diversity in our Twitter study. For instance in source diversity,  Ò$p_i$Ò represents  incoming  tweets  from seed users  with  a specific political  stance,  while $ÓnÓ$ represents all  possible categories. As a result of this formula, the user will have a diversity between 0 and 1, where 0 represents  minimum diversity and 1 represents maximum diversity.  Figure \ref{fig:entropy} shows a user that receives equal amount of tweets (20) from all  political categories and has an incoming diversity of 1. The user only retweets from one political category (10 from Category 1), therefore she/he has an outgoing diversity of 0. 

\subsection{Translating Research Questions into Metrics}
\begin{description}
\item[Source Diversity:] For each user, we used Equation \ref{eq:entropy} to calculate his/her source diversity. For a user A, we compare the tweets published by A's direct followees (people A follows) from different groups of which the political leanings have been categorized as discussed above (See Figure \ref{fig:direct}). This gives us a user's direct input entropy. We then also added the tweets A gets through retweets and investigated if A receives more diverse information through indirect media exposure (See Figure~\ref{fig:indirect}). This provides us a user's indirect input entropy. In both of figures, arrows show the information flow.

\begin{figure}[!tbp]
  \centering
  \begin{minipage}[b]{0.3\textwidth}
    \includegraphics[width=\textwidth]{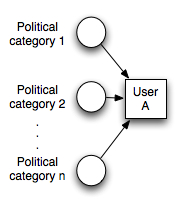}
    \caption{Direct source diversity}
      \label{fig:direct}
  \end{minipage}
  \hfill
  \begin{minipage}[b]{0.3\textwidth}
     \includegraphics[width=\textwidth]{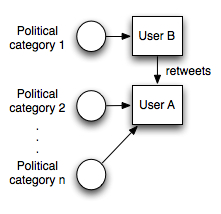}
    \caption{Indirect source diversity}
    \label{fig:indirect}
  \end{minipage}
  
\end{figure}

\begin{figure}[!tbp]
  \centering
  \begin{minipage}[b]{0.38\textwidth}
     \includegraphics[width=\textwidth]{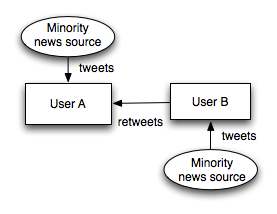}
    \caption{Minority access}
    \label{fig:minority}
  \end{minipage}
  \hfill
    \begin{minipage}[b]{0.38\textwidth}
     \includegraphics[width=\textwidth]{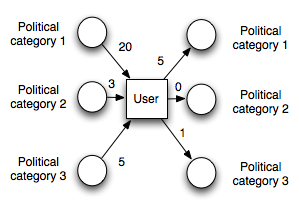}
    \caption{Input-Output Correlation}
    \label{fig:inputoutput}
  \end{minipage}
\end{figure}

\item [Output Diversity:] To measure what the user is sharing after she/he was exposed to different incoming information, we used Equation \ref{eq:entropy} to compare the retweets and replies she/he makes for each political category.
\item [Openness/Minority Diversity:] For this definition of diversity,  we first  defined all seed users who belong to a political party that is either not represented in the parliament, or is represented with few MP's. We also included MP's of a large political party who belong to an ethnic minority. That makes for instance the Kurdish Party BDP and its MP's a minority in Turkey, while we consider the Greens as a minority in the Netherlands. See Appendix A for a list of minorities. Both users defined as minorities create about 15\% of the all observed tweets for both countries.

We then looked whether the user is receiving minority tweets directly or indirectly. For instance, in Figure \ref{fig:minority}, User A is receiving minority tweets by direct subscription, but also indirectly via User B. We defined two metrics to measure minority access. We first look at  the ratio of minority tweets a user gets out of all minority tweets:
\begin{equation}
\label{eq:min-allmin}
\mbox{\# received minority tweets} \over \mbox{\# all published minority tweets}
\end{equation}

We later calculate the ratio of minority tweets in a users' timeline

\begin{equation}
\label{eq:min-all}
\mbox{\# received minority tweets} \over  \mbox{\#received tweets from seeds}
\end{equation}

\item[Input-Output Correlation:] For each user in our sample we look whether the maximum number of  the political position of the messages retweeted by a user is significantly skewed from the political position of the messages that she/he receives. \end{description}
\begin{equation}
\label{eq:inputoutput}
\max(\mbox{incoming political category}) == \max(\mbox{outgoing political category})
\end{equation}

For instance, Figure \ref{fig:inputoutput} shows a  biased user which receives most items from category 1, and also retweets mainly from category 1. 
\section{Results}
This section shows the results for the defined metrics. We tested statistical significance of our results with a two-tailed $t$-Test where the significance level was set to $\alpha = 0.01$ unless otherwise noted. 

\begin{figure}[!tbp]
  \centering
  \begin{minipage}[b]{0.45\textwidth}
    \includegraphics[width=\textwidth]{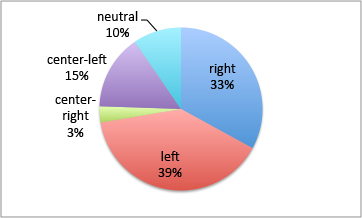}
    \caption{Dutch seed user distribution}
      \label{fig:seed-nl}
  \end{minipage}
  \hfill
  \begin{minipage}[b]{0.41\textwidth}
    \includegraphics[width=\textwidth]{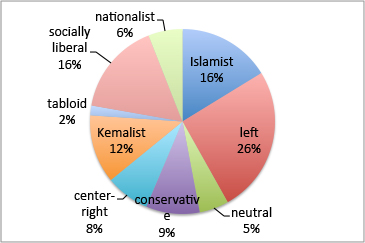}
    \caption{Turkish seed user distribution}
    \label{fig:seed-tr}
  \end{minipage}
  \end{figure}

\begin{figure}[!tbp]
  \centering
  \begin{minipage}[b]{0.45\textwidth}
    \includegraphics[width=\textwidth]{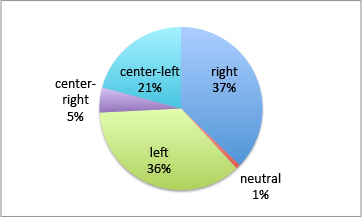}
    \caption{Dutch  user distribution}
      \label{fig:userdistribution_nl}
  \end{minipage}
  \hfill
  \begin{minipage}[b]{0.45\textwidth}
    \includegraphics[width=\textwidth]{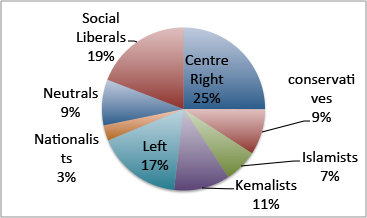}
    \caption{Turkish  user distribution}
    \label{fig:userdistribution_tr}
  \end{minipage}
  \end{figure}

Figures \ref{fig:seed-nl} and \ref{fig:seed-tr} show the distribution of the seed users for both countries. Figures \ref{fig:userdistribution_nl} and \ref{fig:userdistribution_tr} show the distribution of regular users. We see that our selection of popular users covers the political spectrum and it is not concentrated on a single political category.  We have used several sources to do the seed user categorization \citep{Krouwel2008, VanderEijk2000,Trendlight2012,AndreKrouwel, Krouwel2008, Baris2007}.  We  used  the retweet behavior of the users to assign them to a political category to identify their political stance.
  
 \subsection{Seed User Interaction}
To answer research question Q1, Tables \ref{table:seedbias-nl} and \ref{table:seedbias-tr} show the retweet and reply behavior of seed users. Each row shows the category of users who retweet an item or reply to another user. The columns show the source of their retweet or the user they interact with. We observe that 73\% of the left seed users retweet only from left users and reply to left users, while 72\% of the right users do the same. The situation is more extreme for Turkish seed users: 93\% of left seed users only retweet  from and reply to left users, while 94\% of the right seed users show the same behavior.
 
\subsection{Source and Output Diversity}
\begin{figure}[!tbp]
  \centering
  \begin{minipage}[b]{0.49\textwidth}
    \includegraphics[width=\textwidth]{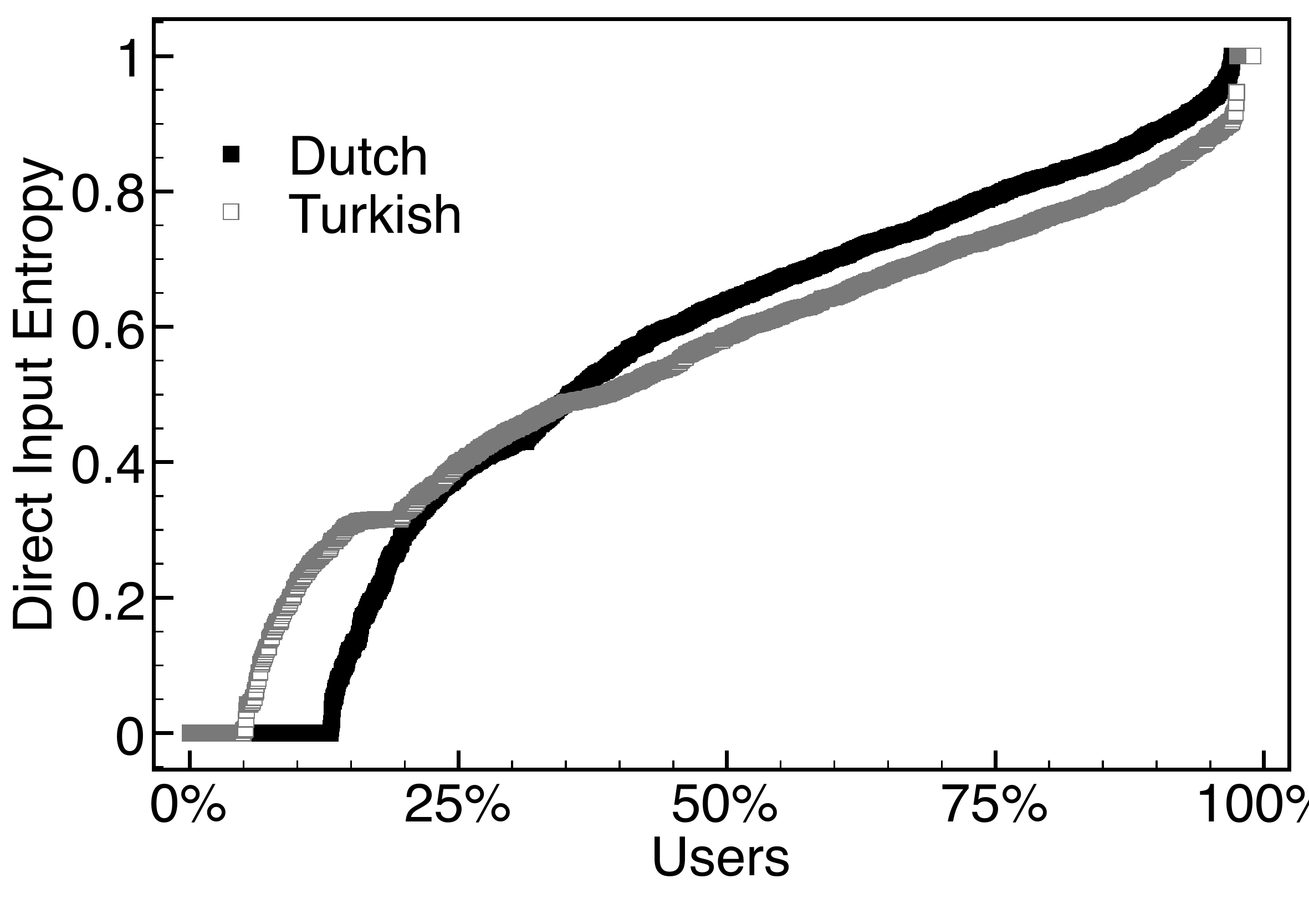}
    \caption{Direct source diversity}
      \label{graph:direct}
  \end{minipage}
  \hfill
  \begin{minipage}[b]{0.49\textwidth}
     \includegraphics[width=\textwidth]{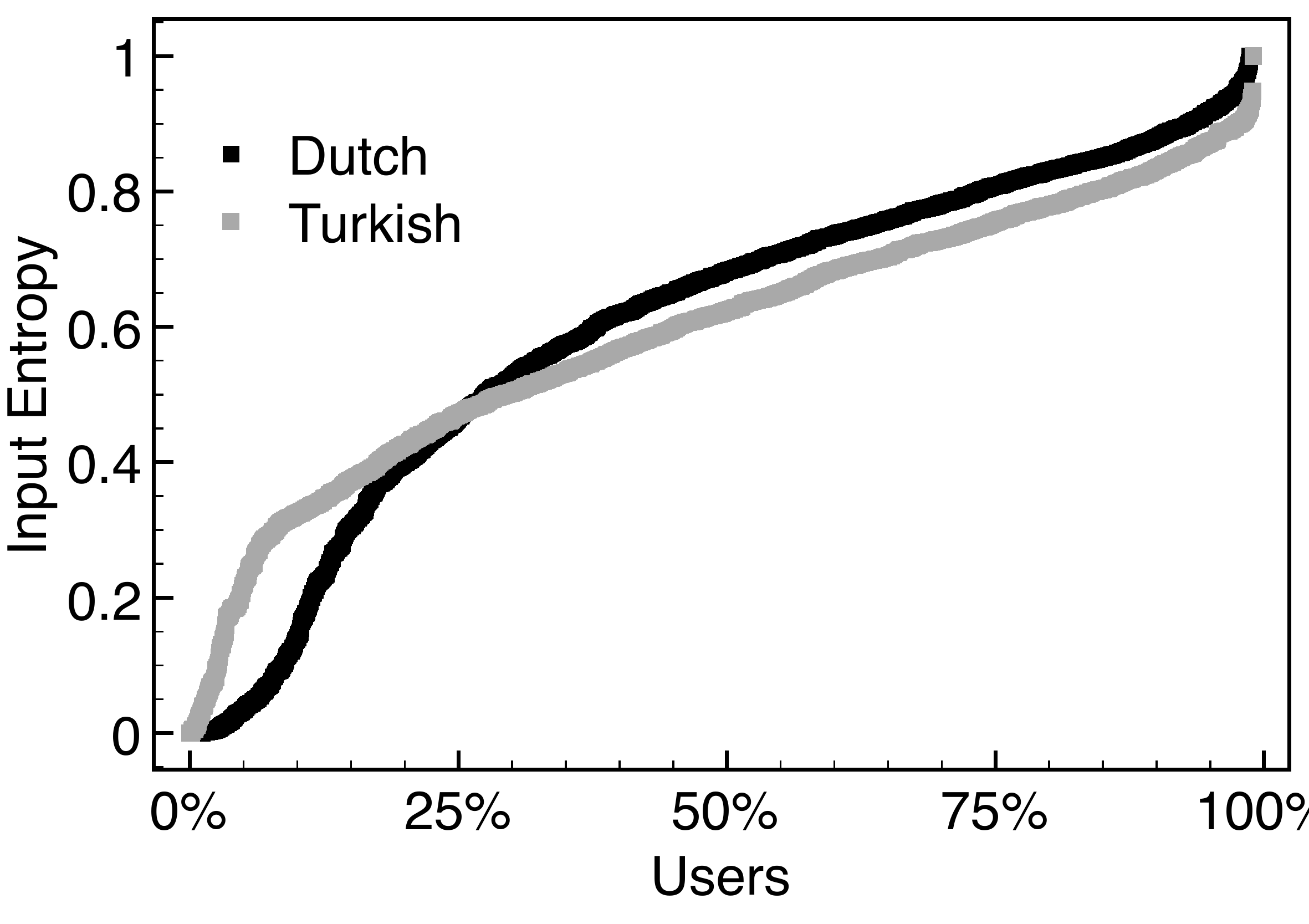}
    \caption{Indirect source diversity}
    \label{graph:indirect}
  \end{minipage}
  
\end{figure}

Table \ref{table:source} shows the results for research questions Q2 and Q3. Here we see that on a scale of 0 to 1, the diversity of the incoming tweets for an average user is approximately 0.6 and the results are not very different for both countries. While diversity is not perfect, we do not observe a true cyberbalkanization and we do not observe a significant difference between two  countries. We observe that indirect communication  (retweets) does increase diversity, but not dramatically.  Figure \ref{graph:direct} and Figure \ref{graph:indirect} show the distribution of source diversity among users. We observe that, indirect communication  decreases the number of users who have a diversity approaching 0 for both countries. Approximately 27\% of the Dutch and 29\% of the Turkish users have an indirect diversity less than 0.5. 

\begin{figure}[!tbp]
  \centering
  \begin{minipage}[b]{0.49\textwidth}
    \includegraphics[width=\textwidth]{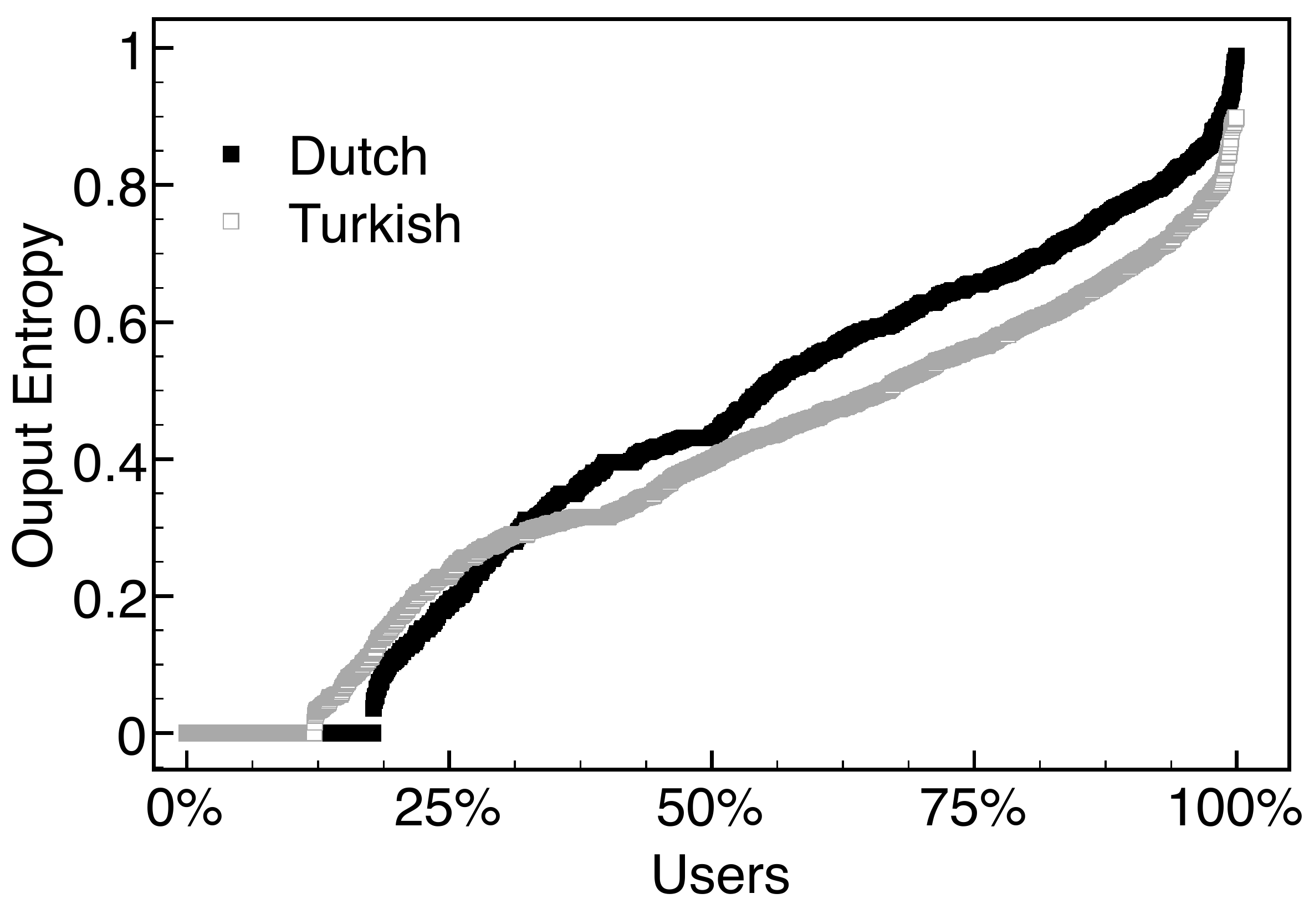}
    \caption{Output diversity (retweet diversity)}
      \label{graph:retweet}
  \end{minipage}
    \hfill
  \begin{minipage}[b]{0.49\textwidth}
    \includegraphics[width=\textwidth]{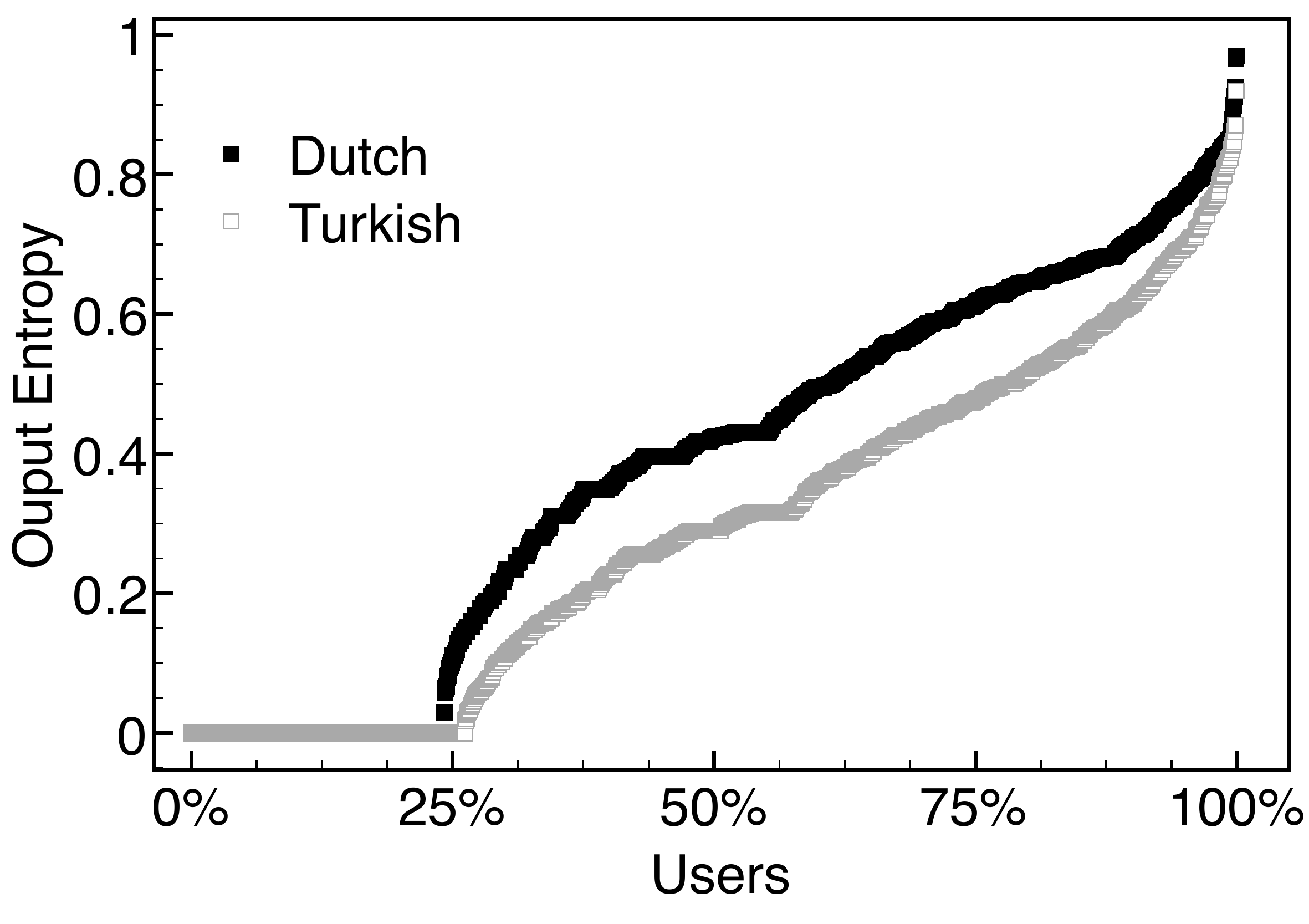}
    \caption{Output diversity (reply diversity)}
      \label{graph:reply}
  \end{minipage}
\end{figure}

However, if we look at the diversity of an average user's output, we see much lower numbers. As Table \ref{table:output} shows,  on a scale of 0 to 1, retweet diversity is 0.43 for the Dutch and 0.40 for the Turkish users. If we look at reply diversity, it is 0.42 for the Dutch  and 0.29 for the Turkish users. Figures \ref{graph:retweet} and \ref{graph:reply} show the distribution of output (retweet and reply) diversity among the population. About 55\% of the Dutch and 66\% of the Turkish users have a retweet diversity lower than 0.5, and about 17\% of the Dutch and 12\% of the Turkish users have a retweet diversity lower than 0.01 ($p<0.001$). About 61\% of the Dutch and 77\% of the Turkish users have a reply diversity lower than 0.5, and about 24\% of the Dutch and 26\% of the Turkish users have a reply diversity lower than 0.01 ($p<0.001$). This means that the users show bias for both their retweet and reply preferences and especially the Turkish reply diversity is quite low.

\subsection{Opennes/Minority Diversity}
\begin{figure}[!tbp]
  \centering
  \begin{minipage}[b]{0.49\textwidth}
    \includegraphics[width=\textwidth]{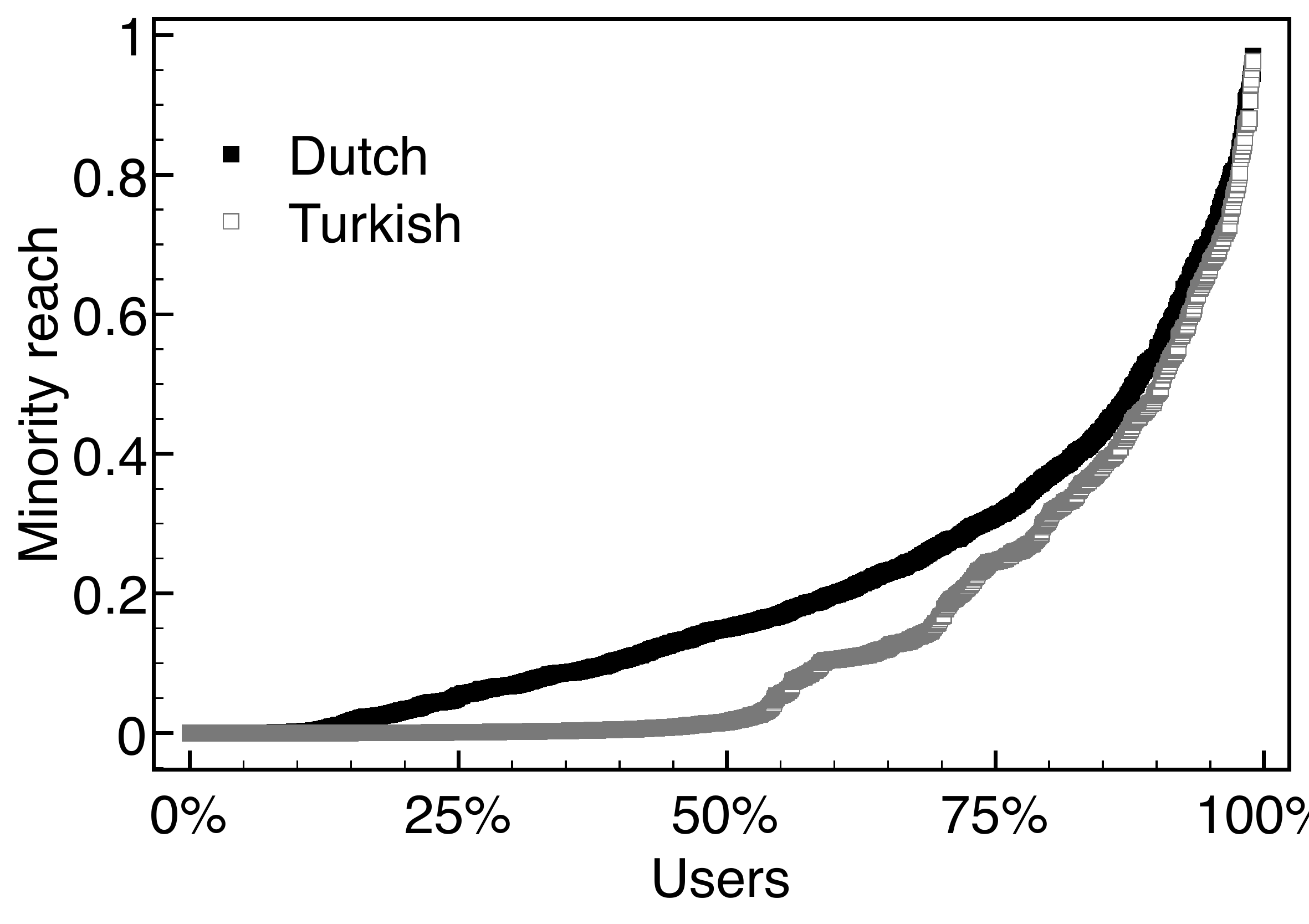}
    \caption{Minority Reach}
      \label{graph:minorityreach}
  \end{minipage}
  \hfill
  \begin{minipage}[b]{0.49\textwidth}
     \includegraphics[width=\textwidth]{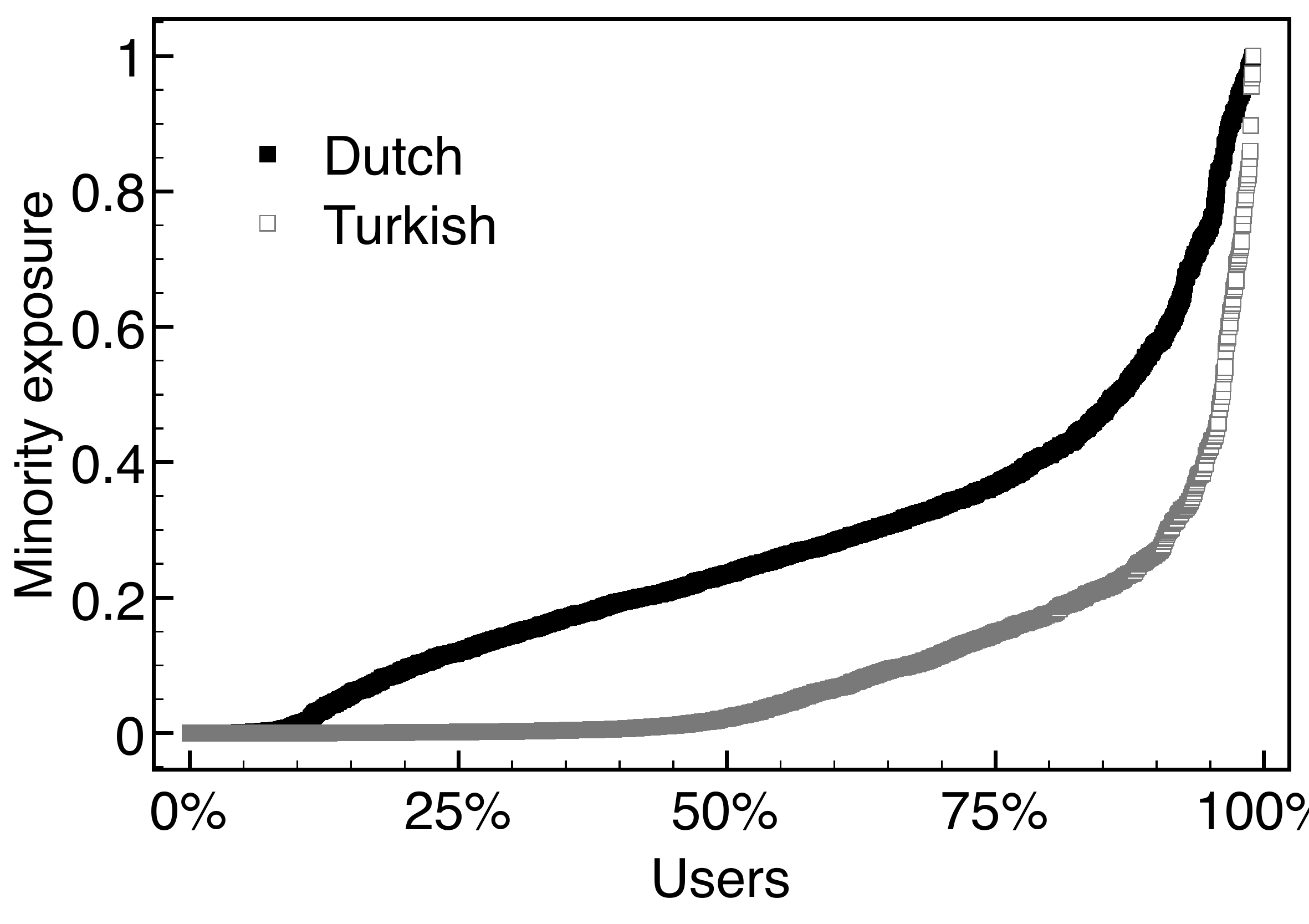}
    \caption{Minority Exposure}
    \label{graph:minorityexposure}
  \end{minipage}
  
\end{figure}

Table \ref{table:minority} shows the results for the research question Q4. First row, which we call ``minority reach" shows the result for Equation \ref {eq:min-allmin} and the second row, which we call ``minority exposure" shows the result for Equation \ref{eq:min-all}.  Minority reach measures how many percent of all the produced tweets by minorities reach an average user. Minority exposure checks the ratio of minority tweets in a user's newsfeed. We observe that an average Dutch Twitter users will receive 15\% of the produced minority tweets, whereas an average Turkish user will only receive 2\% of them. Later, we observe that minority tweets make up 23\% of an average Dutch users' incoming tweets from seed users,  while it only makes up 2\% for a Turkish user. Figures \ref{graph:minorityreach} and \ref{graph:minorityexposure} show the distribution of users for this metric. Here we observe a significant difference between two countries ($p<0.001$). About 55\% of the Turkish users have a minority exposure under 0.05 and 57\% of them have a minority reach under 0.05. The percentages are much lower for the Dutch users: 14\% and 23\% respectively. This means that more than half of the Turkish users are missing almost all the updates produced  by the minorities and their newsfeed contains almost no minority tweets at all. This includes indirect minority tweets thanks to retweets done by their friends.

 \subsection{Input-Output Correlation}
Table \ref{table:userbias} shows the results for the research question Q5. The first row shows the number of users whose output correlates with their input. Such users make up 33\% of the Dutch and 47\% of the Turkish userbase. Further, if we only consider a bias towards a certain political category that is higher than 15\% (for both input and output), 26\% of the Dutch and 36\% of the Turkish users show this behavior.

\begin{table}
\caption{Seed user bias}
\centering
\subfloat[Netherlands]{\label{table:seedbias-nl}\begin{tabular}{ |l|l|l| }
  \hline
User / Source & Left & Right \\ \hline
  Left & 73\% & 27\% \\ \hline
  Right & 28\% & 72\% \\ \hline
\end{tabular}}
\qquad\qquad
\subfloat[Turkey]{\label{table:seedbias-tr}
\begin{tabular}{ |l|l|l| }
 \hline
  User / Source& Left & Right \\ \hline
  Left & 93\% & 7\% \\ \hline
  Right & 6\% & 94\% \\ \hline
\end{tabular}}
\end{table}

\begin{table}
\caption{Different dimensions of diversity. NL = the Netherlands, TR = Turkey.}
\centering
\subfloat[Source Diversity (on a scale of 0 to 1)]{\label{table:source}\begin{tabular}{ |l|l|l| }
  \hline
  & NL & TR \\ \hline
  Direct & 0.63 & 0.58 \\ \hline
  Indirect & 0.68 & 0.62 \\ \hline

\end{tabular}}
\qquad\qquad 
 \subfloat[Output Diversity (on a scale of 0 to 1)]{\label{table:output}
\begin{tabular}{ |l|l|l| }
  \hline
  & NL & TR \\ \hline
 Retweet & 0.43 & 0.40  \\ \hline
 Reply & 0.42 & 0.29  \\ \hline

\end{tabular}}

\subfloat[Input-Output Correlation]{\label{table:userbias}
\begin{tabular}{ |l|l|l| }
  \hline
  & NL & TR \\ \hline
  \# users  & 657 & 828 \\ \hline
  \% users & 33\% & 47\% \\ \hline
\end{tabular}}

\subfloat[Openness/Minority Diversity]{\label{table:minority}
\begin{tabular}{ |l|l|l| }
  \hline
  & NL & TR \\ \hline
  minority reach& 15\% & 2\% \\ \hline
  minority exposure  & 23\% & 2\% \\ \hline
  \%  users under $<$0.05 reach  & 14\% & 57\% \\ \hline
    \%  users under $<$0.05 exposure  & 23\% & 55\% \\ \hline
\end{tabular}}

\end{table}

\section{Discussion}	
\label{sec:discussion}
In this study we have shown different dimensions of diversity and discussed an additional dimension, namely minority access. This dimension has not been previously been subjected to large-scale quantitative validation.  However, as many communication scholars and philosophers have argued, while the media should reflect the preferences present in the society, it should also allow equal access to everyone, including those whose common social location tends to exclude them from political participation. Public life needs to include differently situated voices, not just the majority.

We have shown that different definitions of diversity can be operationalized in different metrics and the question whether ``the filter bubble exists in social media" will have different answers depending on the metric and political segregation of the observed country. For instance, according to the results of our study, source diversity does not differ much for Turkish and Dutch users and we certainly cannot observe a bubble. However, if we consider output, then we  see that the diversity is much lower. Further, if we consider the minority access as a diversity metric, we see that  minorities cannot reach a large percentage of the Turkish population. 

Twitter, with its 140 character limitation, is not the ideal platform for deliberation. However, having a diverse information stream in Twitter is still important, as it can serve as an input for deliberation elsewhere. Information intermediaries such as Twitter could have a considerable influence in nudging people towards more valuable and diverse choices \citep{Helberger2011}. Media diversity has been an important policy objective for the regulation of traditional media \citep{VanCuilenburg2003}. Journalism ethics requires the newspapers to have a balanced and fair coverage of news and opinions and editors and journalists to minimize bias in their filtering decisions. In the abundance of digital online information  and  algorithmic filters to deal with information overload, bias should also be minimized and ideas and opinions of minorities should not be lost.  

Design choices in software codes and other forms of information politics still largely determine the way information is made available and who can speak to whom under what condition \citep{Karppinen2009}. According to Karppinen \citeyearpar{Karppinen2009}, it is  important to make  decisions about standards, because those ``can have lasting influence on media pluralism, even if they are not necessarily  recognized as sites of media policy as such". However, making minority voices reach a wider public is no easy matter. While identifying minorities and their valuable tweets is no easy task, showing these items to ``challenge averse" users is a real challenge \citep{Munson2010}.  For instance Munson et al.\citeyearpar{Munson2013} provided people with feedback  about the political lean of their reading behaviors and found that such feedback had only a small effect on nudging people to read more diversely. More research is needed to understand how users' reading behavior change and to determine the conditions that would allow such a change. 

Further, normative questions arise while making design decisions.  When designing diverse recommendation systems, it is definitely a challenge to determine which view is ``valid". For instance, should a recommendation system show all viewpoints in the Òclimate changeÓ debate, if some viewpoints are not empirically validated or simply seen as false by the majority of the experts? Should a viewpoint get equal attention even if it provides no information or only contains arguments with fallacies? These questions would need to be addressed by a good ethical analysis before design decisions of such systems are made.

\section{Limitations}
This study has several limitations.  First of all, next to the accounts of  traditional media outlets on Twitter, we also selected politicians and bloggers.  While they mainly tweet political matters, it is possible that they have shared personal and non political matters as well. 

Second, while the results give us an idea on the political landscape of the studied countries, Twitter does not represent `all people'. As boyd and Crawford \citeyearpar{boyd2011} have stated,  ``many journalists and researchers refer to `people' and `Twitter users' as synonymous (...) Some users have multiple accounts. Some accounts are used by multiple people. Some people never establish an account, and simply access Twitter via the web". Therefore we cannot conclude that our sample represent the real population of the studied countries. 

Third, input-output correlation does not always implicate that the volume of the content affects the items users share. Users might already be biased before they select their sources and can therefore follow more from certain sources and share from certain categories. 

Fourth, users will make different uses of Twitter. Some might use it as its primary news source, therefore following mainstream items, while others will use it to be informed of the opposing political view or to find items missing in the traditional media. Therefore, we do not know why some users only follow sources from a specific political category. More qualitative studies are needed.

Fifth, retweets in Twitter can be made for different purposes. There is a difference between endorsement retweets (created by pushing the retweet button) and informal tweets (where users include most of the same text often prefixed by `RT' or similar but also add their own comments before or after the tweet). These two actions measure a different interaction. Informal retweets and replies could also express disagreement and show us deliberation. In order to make a distinction between these two types of tweets, we need semantic analysis. We are not aware of the availability of such tools for the Turkish language, therefore we were unable to perform such an analysis. While information diversity is important not only for  deliberative models of democracy, it would be very useful to study deliberation on Twitter by using such tools in the future. 

Sixth, users could retweet or reply with bad intentions, such as trolling. For retweets, we only measured users' retweets to original tweets created by seed users. We assume that, those powerful political actors would not take part in trolling. Users can retweet a seed user's tweet randomly or for trolling purposes. Same issue can manifest itself in replies. Since we did not perform a semantic analysis, this remains a limitation.

\section{Conclusion and Future Work}
In this paper, we have introduced a framework that lists metrics used in the previous social media analytics studies and added new ones using the theory from other fields. As one of the outcomes of the results from the previous section, we showed how minority access is missing from the typical dimensions of viewpoint diversity studied by computer scientists and the impact it has on viewpoint diversity analysis. To our knowledge, this is the first work that provides an overview of all the used metrics in social media analytics literature and the first study to apply an ``openness" metric. Building on this framework, new studies can be performed for different countries and political cultures. For instance, Belgium, a country where different languages are spoken in different regions is an an interesting case to apply our framework.

In the recent months, Turkey experienced several political protests that spontaneously erupted against the destruction of trees and the building of a shopping mall at Gezi Park in Taksim Square and large scale corruptions within the government. Twitter and Facebook played a vital role during these movements and became the only communication medium when traditional media performed self-censorship \citep{Dorsey2013,hammond2013,Oktem2013}. It would be very useful to see whether the political stance of our observed users have changed. It is also challenging to identify the opinion leaders during these movements and find whether they communicate with each other or form their own ``bubbles". It is further valuable to see if minorities were able to reach a wider public during those protests. A hashtag based political communication or an extension of our methodology could bring new insights. 

Our study was focused on Twitter and studied whether users have put themselves in bubbles by following  individuals from only one end of the political spectrum and showed a biased sharing behavior. Twitter itself does not employ a personalization algorithm in a user's timeline. However other social networking platforms, such as Facebook, do use a personalization algorithm and filter certain information on user's behalf \citep{Bozdag2013}. Future studies can perform black-box testing techniques to determine whether filters used by these platforms lead to bubbles (See \cite{Jurgens2013}). Creating multiple profiles while modifying certain factors, such as political affiliation, age, location, etc. can help us detect bubbles, if they exist.

\section*{Appendix A: List of Minorities}
\label{app:minorities}
Dutch minorities: Keklik Yucel, SGP, Khadija Arib, Vera Bergkamp, Sadet Karabulut, Farshad Bashir, Tanja Jadnanansing, Piratenpartij NLD, Partij van de Dieren, Fatma Koser Kaya, ChristenUnie, Groenlinks,  Marianne Thieme, Femke Halsema.

Turkish minorities: Ayca Soylemez, Evrensel, Aydinlik, Ozgur Gundem, Pinar Ogunc, Bianet, Sebahat Tuncel, Sol Haber Portali, Halkin Gazetesi Birgun
 Yildirim Turker, BDPGenelMerkez, Ufuk Uras, Selahattin Demirtas, Sirri Sureyya Onder, Sinan Ogan, Hasip Kaplan.

Note that both minorities create about 15\% of all tweets produced by seed users.

\bibliographystyle{model5-names}
\bibliography{library}

\end{document}